\documentclass[aps,pra,10pt,twocolumn,superscriptaddress,nofootinbib]{revtex4-2}

\usepackage[english]{babel}
\usepackage[utf8]{inputenc}
\usepackage{amsmath,amssymb,amsfonts,dsfont,graphicx,xcolor,mathtools,dsfont,bm,hyperref,physics}
\usepackage[outline]{contour}
\usepackage[normalem]{ulem} 
\hypersetup{
	colorlinks=true,
	linkcolor={red!40!black},
	citecolor={blue!60!black},
	urlcolor={blue!50!black}
	}
\definecolor{myred}{rgb}{0.85,0,0}
\definecolor{mygray}{rgb}{0.87, 0.87, 0.87}

\let\oldbibitem\bibitem 
\renewcommand{\bibitem}{
    \renewcommand{\doi}[1]{\texttt{\href{https://doi.org/##1}{doi:##1}}} 
    \let\bibitem\oldbibitem 
    \oldbibitem 
}

\begin{document}

\title{Purified phase estimation samples spectra efficiently}

\affiliation{Fujitsu Research of Europe Ltd., SL1 2BE Slough, U.K.\\
${}^{2}$University of Exeter, Department of Physics and Astronomy, Stocker Road, Exeter EX4 4QL, UK\\
${}^{3}$Fujitsu Research of Europe Ltd., Pozuelo de Alarcón, 28224 Madrid, Spain}

\author{Stefano Scali\textsuperscript{\textpilcrow,1,2}}
\email{stefano.scali@fujitsu.com}
\email{s.scali@exeter.ac.uk}

\author{Josh Kirsopp\textsuperscript{\textpilcrow,1}}
\email{josh.kirsopp@fujitsu.com}

\author{Antonio Márquez Romero\textsuperscript{3}}
\email{antonio.marquezromero@fujitsu.com}

\author{Michał Krompiec\textsuperscript{1}}
\email{michal.krompiec@fujitsu.com}

\begin{abstract}
Quantum phase estimation (QPE) is a cornerstone algorithm for extracting Hamiltonian eigenvalues, but its standard, eigenstate-centric form relies on carefully prepared coherent inputs that are costly or impractical for many strongly correlated systems. We overcome this bottleneck via DOS-QPE, an incoherent, purification-based variant of QPE that works directly with mixed-state probes and estimates the density of states (DOS) of the Hamiltonian. By adding a purification register and simple entangling layers, we turn standard QPE into an ensemble-based DOS sampler without modifying the core phase-estimation block. Conceptually, this purification closely aligns with the recent random purification channel framework from quantum learning theory, but instantiated here as a concrete phase-estimation circuit. We further equip DOS-QPE with symmetry-adapted input ensembles and a compressed-sensing reconstruction pipeline, and demonstrate on fermionic and nuclear Hamiltonians that a single experimental setup can recover rich spectral information relevant to thermodynamics, spectroscopy, and many-body structure.
\end{abstract}

\maketitle
\renewcommand{\thefootnote}{\fnsymbol{footnote}}
\footnotetext[5]{These authors contributed equally to this work.}
\renewcommand{\thefootnote}{\roman{footnote}}


\section{Introduction}
\label{sec:intro}

Quantum phase estimation (QPE) remains the gold standard for extracting eigenvalue information in the form of quantum phases~\cite{kitaev1995quantum,Abrams_1997,Abrams_1999,dobvsivcek2007arbitrary}. In its textbook formulation, QPE is an eigenstate-centric protocol: one supplies a carefully prepared pure state with large overlap on a target eigenvector, and the circuit returns an estimate of the associated phase. For strongly correlated or structurally complex Hamiltonians, however, preparing such trial states may require deep, problem-specific circuits or demanding variational algorithms~\cite{plesch2011quantum,sun2023asymptotically,zhang2022quantum}. On real devices, noise and decoherence further erode the assumption that a single, well-isolated eigenstate can be reliably prepared. This motivates a different question: \emph{can we use the machinery of phase estimation to interrogate entire spectral subspaces via mixed or ensemble states instead of finely tuned eigenstates?}

Purification offers a natural route to such an “incoherent” formulation of QPE. Any mixed state can be realized as the reduced state of a larger pure state on a system plus auxiliary register~\cite{Nielsen_2012,Chiribella_2010}. Conceptually, nothing prevents us from feeding this purification into a QPE circuit and tracing out the purifying system at the end. Despite the ubiquity of purifications in quantum information, this perspective has not been developed as a general circuit primitive for phase estimation. Existing uses are scattered across specific settings, such as \emph{garbage states} in quantum topological data analysis~\cite{Lloyd2016} or maximally mixed probes for Hamiltonian moments in signed graphs~\cite{scali2025signed}, rather than forming a unified framework for ensemble-based spectral extraction.

In this work we take purification as the organizing principle for QPE and build a density-of-states variant, DOS-QPE, around it. By adding a dedicated purification register and preparing controlled entanglement between the system and its purifier, we turn standard QPE into an incoherent primitive that samples the density of states of the Hamiltonian generator. The QPE block itself remains essentially unchanged; what we redesign is the input. Instead of a single pure eigenstate, we supply mixed or symmetry-adapted ensembles implemented via simple purification circuits. This shift allows us to access thermodynamic and spectral information directly from ensembles, and to probe entire invariant subspaces without ever preparing individual eigenstates.

While developing this purified-QPE framework, we became aware of concurrent progress in quantum learning theory that points in a similar direction. The line of work by Tang et al.~\cite{tang} and Pelecanos et al.~\cite{pelecanos} shows that mixed-state learning and tomography can be efficiently reduced to pure-state problems via a \emph{random purification channel} (RPC), an instance of the “acorn trick’’ converting multiple copies of a mixed state into random purifications suitable for pure-state algorithms. Girardi et al.~\cite{girardi} further provide an explicit and simple construction of the RPC. These results offer a natural theoretical backdrop: they identify purification not as a technical afterthought, but as a fundamental algorithmic resource. In hindsight, DOS-QPE can be viewed as the phase-estimation analogue of this sample-efficient paradigm, leveraging purification to turn incoherent ensembles into usable input for a standard, well-studied quantum subroutine.

We deliberately base our DOS estimation framework on QPE rather than on alternative approaches such as Chebyshev-moment methods~\cite{Summer_2024}. QPE sits at the core of many quantum algorithms, and recasting it in an incoherent, purification-based form preserves compatibility with this broader ecosystem. Our primitive can in principle be used as a drop-in component in QPE-based routines, including Shor’s algorithm~\cite{Shor}, quantum simulation for quantum chemistry~\cite{Aspuru_Guzik_2005}, and the HHL algorithm for solving linear systems~\cite{Harrow_2009}, enabling ensemble-based versions of otherwise eigenstate-driven procedures.

In this work, we generalize and refine a QPE-based method for DOS estimation (DOS-QPE)~\cite{scali2025signed}, and elevate it to a reusable circuit primitive. We (i) formalize DOS-QPE in a purified setting, making explicit how a purification register and simple entangling circuits turn standard QPE into an incoherent DOS sampler; (ii) introduce mixed-state probes tailored to specific symmetry sectors, including maximally mixed ensembles and particle-number–resolved probes constructed from Dicke states; and (iii) develop a convex-optimization–based pipeline for spectrum reconstruction using quadratic programming and compressed sensing~\cite{Cornu_jols_2018,Donoho_2006,Cand_s_2006}. We then demonstrate the reach of this approach on three representative classes of Hamiltonians: a Fermi–Hubbard model, an electronic-structure Hamiltonian, and a strongly correlated nuclear Hamiltonian. Together, these examples showcase DOS-QPE as a practical and versatile primitive for spectral subspace extraction, highlighting its potential as a core tool for early fault-tolerant quantum simulations.


\section{DOS-QPE}
\label{sec:dosqpe}
\begin{figure*}
    \centering
    \includegraphics[width=\textwidth]{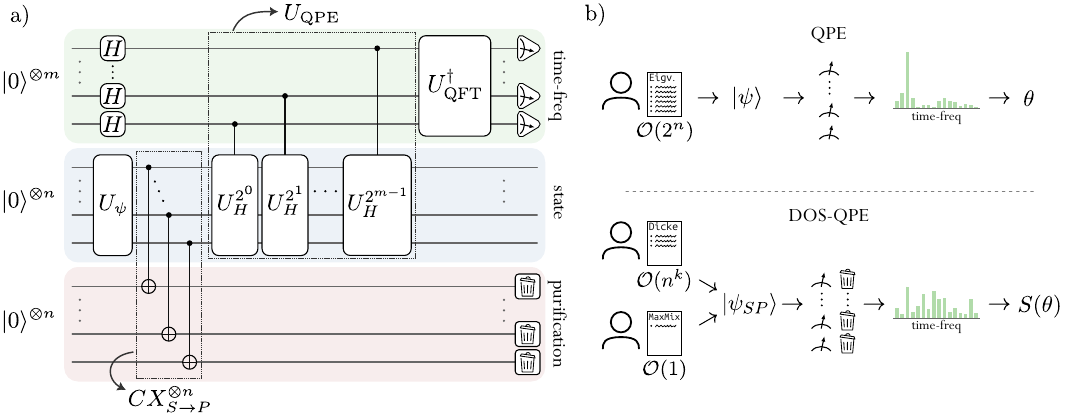}
    \caption{\textbf{a) DOS-QPE circuit primitive.} In panel a), we show the DOS-QPE as a circuit primitive. The full circuit consists of three registers: a time-frequency register, a state register, and a purification register. The additional purification register is maximally entangled to the state register to create the state ensemble. This is done through a cascade of serialized CNOT operations, $CX^{\otimes n}_{S\rightarrow P}$. This last register is traced out at the end of the protocol. \textbf{b) QPE v. DOS-QPE algorithmic comparison.} In panel b), we compare the standard QPE algorithm to the DOS-QPE algorithm. While QPE relies on the preparation of good-enough states $|\psi\rangle$  with sufficient overlap with the target eigenstates, DOS-QPE allows for probing entire subspaces via purifications $|\psi_{SP}\rangle$ of simple construction. This translates to a cheaper state preparation routine. After measuring the time-frequency register in QPE, one estimates the phase $\theta$. In DOS-QPE, we measure the time-frequency register and trace out the purification register to estimate the eigenvalue function $S(\theta)$.}
    \label{fig:dosqpe}
\end{figure*}

Standard quantum phase estimation consists of a state register of dimension $n$ and a time-frequency register of dimension $m$, often referred to as the ancilla register. In QPE, we are interested in the extraction of the phase $\theta$ relative to an eigenstate $|\psi\rangle$ of a unitary operator ${U}_H$, where ${U}_H|\psi\rangle = e^{2\pi i \theta}|\psi\rangle$. This can be any unitary $n$-qubit gate, often generated by the Hamiltonian ${H}$ as $U_H = e^{-iHt}$. To achieve this, we prepare the state register with the eigenstate $|\psi\rangle$ and perform the controlled-unitary operations ${U}_\text{C} = \sum_{k=0}^{2^m-1}|k\rangle\langle k| \otimes {U}_H^k$. This cascade of unitary operations acting onto the state register at incremental ``times'' $k$ extracts the phase via phase kickbacks. The simplest example of this process is the Hadamard test, formally a QPE circuit with a single-qubit time-frequency register. These phases are encoded into the time-frequency register and can be accessed in the form of probabilities after performing an inverse quantum Fourier transform (iQFT), ${U}_\text{QFT}^\dagger$, on it. In Fig.~\ref{fig:dosqpe}a), the QPE circuit would correspond to the first two registers, namely the ``time-frequency'' and the ``state'' registers, thus removing the ``purification'' register and relative operations.
Standard QPE is the fault-tolerant benchmark for the estimation of phases (eigenvalues) in the case of a pure, coherent eigenstate of the unitary~\footnote{Throughout this paper, we refer to coherent and incoherent in the sense of resource theories~\cite{aberg, Baumgratz_2014}. Specifically, we refer to states as coherent when their density matrix contains non-zero off-diagonal elements (coherences), and as incoherent when these off-diagonal terms vanish.}. Its successful execution requires the prepared probing state to have substantial overlap with the target eigenstate whose phase we aim to measure. Although several techniques have been proposed to address the state preparation problem, this step often incurs significant overhead and is sometimes the most challenging part of the protocol.

In chemistry, trial eigenstate preparation remains a hot topic~\cite{lee2023evaluating}; for weakly correlated systems, the qubit-mapped Hartree-Fock state may be good enough~\cite{tubman2018postponing}, but for more strongly correlated systems, elaborate state preparation protocols are required. These usually take the form of a variational quantum algorithm~\cite{peruzzo2014variational}, which represent difficult optimization problems~\cite{larocca2024review, astrakhantsev2023phenomenological} and suffer prohibitive sampling costs~\cite{gonthier2022measurements}. Otherwise, approximate classical eigenstates can be loaded onto a qubit register by an appropriate mapping such as CVO-QRAM~\cite{deveras2022double}. Still, even this approach does not scale favorably with system size, not to mention the additional classical cost of finding a high-quality approximation to the target eigenstate, using, for example, one of the selected CI methods~\cite{huron1973iterative,tubman2016deterministic,sharma2017semistochastic}. In addition, we might be interested in more general properties of the Hermitian operator generator of the unitary evolution. For example, accessing information encoded in higher-energy states may require reconstructing the entire spectrum of the target Hamiltonian. This is the case we will consider in the following.

We are interested in the reconstruction of the so-called eigenvalue function~\cite{osborne2006renormalisationgroup} or density of states of the Hamiltonian operator ${H}$. The density of states (DOS) plays a central role in extracting spectral properties of an operator. It is typically derived from the Fourier transform of the operator's unitary time evolution and is essential for the reconstruction of the microcanonical ensemble of the system~\cite{kardar}. The DOS technique is a flexible approach for gathering information about both low- and high-frequency components.

Consider a Hermitian operator ${H}$ and its spectral resolution, ${H} = \sum_k \theta_k |k\rangle\langle k|$, where $\theta_k$ are the eigenvalues corresponding to the eigenvectors $|k\rangle$. The DOS is the eigenvalue distribution of the operator ${H}$,
\begin{equation}
    \label{eq:dos}
    S(\theta) = \sum_k\delta(\theta - \theta_k) .
\end{equation}
Looking at the spectral decomposition of the Hamiltonian ${H}$, that is, an ensemble of pure states $|k\rangle\langle k|$, one might ask if we can extract the DOS function from the standard routine of phase estimation. This can indeed be done using the purification theorem. Purification is a fundamental concept of quantum information theory~\cite{Nielsen_2012} that expresses any mixed state as the reduced state of a larger pure state having a specific entanglement structure, i.e., that of a maximally entangled state. For any mixed state $\rho_S$ on a finite-dimensional Hilbert space $\mathcal{H}_S$, there exists a pure state $\ket{\psi}_{SP}$ in a larger Hilbert space $\mathcal{H}_S \otimes \mathcal{H}_P$ such that
\begin{equation}
    \label{eq:purification}
    \rho_S = \Tr_P \left( \dyad{\psi}_{SP} \right) .
\end{equation}
Here, $\mathcal{H}_P$ is an auxiliary system (often called the \emph{purifying system}) with dimension at least equal to $\text{rank}(\rho_S)$. For simplicity and ease of explanation, we fix the purifying system to the same dimension as the state register.

To implement this approach into standard QPE (see App.~\ref{sec:qpe_appendix} for the derivation), we need to add an additional ancillary register that we will refer to as the purification register. This register matches the state register dimension, but it will be traced out at the end of the protocol. By means of this operation, we are effectively evaluating the trace over the system $P$ in Eq.~\eqref{eq:purification}. This will allow us to translate QPE from a coherent to an incoherent approach, extending the routine to support further practical applications. Note that purification is not the only way to transition between coherent and incoherent QPE. In case of interest in statistical properties of the systems, any thermal state preparation routine~\cite{gilyen,Ding_2024,Ding_2024_2} will suffice as long as the output state can be matched to the state register.

As a result of the purification, the measurement outcome $y \in \{0, \dots, 2^m - 1\}$ from the time-frequency register yield coarse-grained phase estimates $\tilde{\phi}_j = y / 2^m$ with probability
\begin{equation}
    P(y) = \sum_j p_j |c_j(y)|^2,
\end{equation}
where $p_j$ are the spectral weights of $\rho_S$ and $|c_j(y)|^2$ encodes the QPE response kernel. This empirical distribution reconstructs the coarse-grained DOS at resolution $\Delta\phi = 1/2^m$. The total reconstruction error scales as
\begin{equation}
    \|\hat{\rho}_S - \rho_S\|_2 \sim \sqrt{ \frac{N_{\mathrm{eff}} \cdot 2^m}{M} },
\end{equation}
where $N_{\text{eff}}$ is the dimension of the subspace covered by the ensemble contributing to $\rho_S$, and $M$ is the number of QPE repetitions. Thus, achieving a target error $\delta$ requires
\begin{equation}
    M \gtrsim \frac{N_{\mathrm{eff}} \cdot 2^m}{\delta^2}.
\end{equation}

For details on the derivation of the standard QPE and DOS-QPE results, including error analysis and scaling, see App.~\ref{sec:qpe_appendix} and App.~\ref{sec:dosqpe_appendix}, respectively.


\subsection{Mixed-state probes}
\label{sec:mixed_states}

In DOS-QPE, the choice of input ensemble determines which part of the spectrum is probed and with what weights. Rather than relying on a single coherent eigenstate, we work with mixed states
\begin{equation}
    \rho_S = \sum_j p_j \dyad{\psi_j},
\end{equation}
whose spectral weights $\{p_j\}$ induce a corresponding weighting of the Hamiltonian eigenvalues in the measured DOS. As discussed in Sec.~\ref{sec:dosqpe}, any such state can be implemented via purification: we embed $\rho_S$ into a larger pure state $\ket{\psi}_{SP}$ on $\mathcal{H}_S \otimes \mathcal{H}_P$ and trace out the purifying system $P$ at the end of the protocol,
\begin{equation}
    \rho_S = \Tr_P \bigl( \dyad{\psi}_{SP} \bigr).
\end{equation}

Operationally, this corresponds to adding an extra purification register, preparing an entangled state between $S$ and $P$ with a simple circuit primitive, and then discarding $P$ after QPE. In this work, we focus on two practically relevant classes of mixed states:
(i) the maximally mixed state, which leads to a flat sampling over the entire Hilbert space, and
(ii) particle-number–resolved mixed states obtained from Dicke-state probes, which restrict DOS-QPE to a fixed Hamming-weight sector.

These ensembles can be realized in two complementary ways. First, we introduce an explicit circuit construction based on an additional purification register and a layer of CNOT gates in cascade that prepares a maximally entangled state between $S$ and $P$, which is then traced out. Second, one may employ the recent Haar-random purification methods developed in the context of quantum learning theory by Tang et al.~\cite{tang}, Pelecanos et al.~\cite{pelecanos}, and Girardi et al.~\cite{girardi}. There, the random purification channel (RPC) efficiently converts multiple copies of a mixed state into Haar-random purifications that can be processed by pure-state algorithms. In our setting, such Haar-random purifications provide an alternative, theoretically grounded route to preparing mixed-state probes compatible with DOS-QPE without tailoring a problem-specific purification circuit.


\subsubsection{Maximally mixed state via Hadamard layer}
\label{sec:maximallymixedstates}

The most standard example of a mixed-state probe is the maximally mixed state on the $n$-qubit state register,
\begin{equation}
    \rho_S = \frac{\mathbb{I}}{2^n},
\end{equation}
which assigns equal weight to all computational basis states and therefore to all eigenstates of the Hamiltonian under a generic encoding. A convenient purification of $\rho_S$ is the maximally entangled state
\begin{equation}
    \label{eq:max_entangled_phi}
    \ket{\Phi}_{SP}
    = \frac{1}{\sqrt{2^n}} \sum_{k=0}^{2^n-1} \ket{k}_S \otimes \ket{k}_P,
\end{equation}
where $\{\ket{k}_P\}$ is a computational basis for the auxiliary system $\mathcal{H}_P \cong \mathcal{H}_S$, and $S$ and $P$ denote the state and purification registers, respectively. Tracing out $P$ yields
\begin{equation}
    \Tr_P\bigl(\dyad{\Phi}_{SP}\bigr) = \frac{\mathbb{I}}{2^n} = \rho_S,
\end{equation}
that is, the maximally mixed state on $\mathcal{H}_S$.

This purification can be implemented by a simple circuit primitive. Starting from $\ket{0}^{\otimes n}_S \otimes \ket{0}^{\otimes n}_P$, we first apply a layer of Hadamard gates $H^{\otimes n}$ on the state register $S$ to prepare a uniform superposition, and then apply a cascade of CNOT gates from $S$ to $P$,
\begin{equation}
    \ket{0}^{\otimes n}_S \ket{0}^{\otimes n}_P
    \xrightarrow{H^{\otimes n} \otimes \mathbb{I}}
    \frac{1}{\sqrt{2^n}} \sum_k \ket{k}_S \ket{0}^{\otimes n}_P
    \xrightarrow{\text{CX cascade}}
    \ket{\Phi}_{SP}.
\end{equation}
After running QPE with controlled-$U_H^k$ acting only on $S$, we trace out (or equivalently, discard) the purification register $P$, thereby realizing DOS-QPE with a maximally mixed probe. Using this ensemble with flat distribution, we can probe all of the Hamiltonian's eigenvalues with equal probability and estimate the Hamiltonian's spectral resolution. The power and potential of maximally mixed states in QPE have been demonstrated under the umbrella of garbage state preparation and kernel extraction in quantum topological data analysis~\cite{Lloyd2016}, and, with the addition of the DOS-QPE circuit primitive, in generating features for learning the frustration index in signed graphs~\cite{scali2025signed}.


\subsubsection{Particle-number–resolved mixed states via Dicke probes}
\label{sec:dicke_states}

In many fermionic simulations, the eigenstates of interest lie within a fixed-particle-number sector. To exploit this structure, we consider mixed-state probes supported only on the subspace of fixed Hamming weight $k$. Let $\mathcal{F}$ be the $2^M$-dimensional Fock space associated with $M$ discrete fermionic modes, which decomposes as
\begin{equation} \label{eq:fock_space_decomposition}
    \mathcal{F}
    = \bigoplus_{k=0}^M \mathcal{F}_k,
\end{equation}
where each $\mathcal{F}_k$ is spanned by occupation number vectors with particle number $k$. A 1-particle state $\ket{x_i}$ is either $\ket{0}$ or $\ket{1}$, corresponding, in second quantization, to an empty or occupied mode. A generic $k$-body basis state can be written as
\begin{equation}
    \ket{\bm{x}} = \bigotimes_{i=1}^M \ket{x_i},
\end{equation}
with Hamming weight $|\bm{x}| = k$.

It is often the case that the physically relevant spectrum is contained in a fixed $\mathcal{F}_k$. If we were to use the maximally mixed state on the full Hilbert space as the DOS-QPE probe, the resulting spectrum may be unnecessarily crowded and incur a large sampling overhead, as many samples would fall outside the sector of interest. A simple way to enforce particle-number symmetry in DOS-QPE is to choose a fermion–qubit encoding that preserves Hamming weight, such as the Jordan–Wigner mapping~\cite{JordanWigner},
\begin{equation}
    \ket{x_0, x_1, \ldots, x_{M-1}}
    \longrightarrow
    \ket{q_0, q_1, \ldots, q_{n-1}},
\end{equation}
with $n = M$ and $x_i = q_i \in \{0,1\}$. Under this encoding, the two projections of each qubit represent empty or occupied fermionic modes.

To restrict the probe to the correct subspace, we prepare an even superposition over all computational basis states of Hamming weight $k$ on the state register before purification. These are the Dicke states,
\begin{equation}
    \ket{D^n_k}
    = \binom{n}{k}^{-1/2}
      \sum_{\substack{x \in \{0,1\}^n \\ |x| = k}} \ket{x},
\end{equation}
where $n$ is the number of qubits and $k$ is the desired Hamming weight. Deterministic circuit constructions for Dicke states are known with depths scaling as
$\mathcal{O}\!\bigl(k \log (n/k)\bigr)$, $\mathcal{O}\!\bigl(k \sqrt{n/k}\bigr)$, and $\mathcal{O}(n)$ on devices with all-to-all, grid, and linear nearest-neighbour connectivities, respectively~\cite{bartschi2022short}. By purifying $\ket{D^n_k}$ using the same CX-cascade entangler between $S$ and $P$ and tracing out $P$ after QPE, we obtain a mixed-state probe supported only on the fixed-$k$ sector.

Perhaps the most natural application of DOS-QPE with a Dicke-state probe is the extraction of eigenspectra of many-body Hamiltonians at fixed particle number. In standard QPE and many of its extensions, one must prepare a trial state with large overlap with a particular eigenstate, often through variational algorithms such as VQE, which are costly to optimize and may suffer from barren plateaus. In contrast, Dicke-probed DOS-QPE does not require preparation of individual fermionic eigenstates, preserves particle-number symmetry under Jordan–Wigner, and allows a single experimental configuration to recover the full eigenspectrum within the chosen particle-number sector. The dominant classical cost in the simplest workflow is that of a single Hartree–Fock calculation, which scales polynomially with system size even for naive implementations, plus the cost of constructing the corresponding quantum circuit.


\section{Spectrum reconstruction}
\label{sec:spectrumreconstruction}
Transitioning from estimating a single phase using QPE to reconstructing the full spectrum with DOS-QPE requires a more refined post-processing step. This additional cost is justified by the richer information retrieved from the spectrum. DOS-QPE effectively samples from the normalized density of states, 
\begin{equation}
    S(\theta) = \frac{1}{2^n} \sum_{i=1}^{2^n} \delta(\theta - \theta_i) ,
\end{equation}
where, as a refinement of Sec.~\ref{sec:dosqpe}, we have included the normalization factor. Here, the eigenvalues $\theta_i$ of $H$ are each repeated with their multiplicity. In practice, with only $m$ qubits in the time-frequency register, one obtains a discretized histogram
\begin{equation}
    P_i = \int_{I_i} S(\theta)d\theta + \text{(noise)} , \quad i=0,1,\dots, 2^m-1 ,
\end{equation}
where the $I_i$ are adjacent intervals of width $\Delta\theta = 2\pi/2^m$.

The task of spectrum reconstruction is to recover both the continuous eigenvalues locations ${\theta_i} \subset \mathbb{R}$ and their degeneracies ${d_i} \subset \mathbb{Z}_{\geq 0}$ from the sample histogram $P=\{P_i\}$. This problem is challenging for several reasons: (i) finite resolution due to limited register size $m$, which determines the bin width $\Delta\theta$; (ii) noise, including statistical fluctuations of order $\mathcal{O}(n^{-1/2})$; (iii) unresolved spectral gaps and an unknown number of distinct eigenvalues. While some of these challenges will result in inevitable inaccuracy of the reconstructed signal, we will resort to techniques such as compressed sensing and convex optimization to efficiently find solutions.


\subsection{The degeneracy integer problem}
\label{sec:degeneracy}
With access to the readout of the time-frequency register of the DOS-QPE circuit, we aim to estimate the spectral resolution, ${H} = \sum_k \theta_k |k\rangle\langle k|$. To do so, we face the problem of generating the best spectral estimate out of the readout probability distribution $P$. We refer to this estimation task as the degeneracy integer problem, falling under the umbrella of spectral density estimation~\cite{stoica2005spectral}.

The problem consists of assigning a set of normalized degeneracies $\{\tilde{d}_i\}_i$, subject to $\sum_i \tilde{d}_i = \sum_i d_i / \dim({H}) = 1$, to minimize a suitable distance $D(P, \tilde{P})$ between the target probability distribution $P$ and the reconstructed spectrum $\tilde{P}$. While the ideal reconstruction would be $P(\theta) = \sum_i d_i \delta(\theta-\theta_i)$, the smeared, finite measured distribution will be a convolved distribution $\tilde{P}(\theta_j) = \sum_i d_i K(\theta_j - \theta_i)$, where $K$ is the kernel and $\theta_j$ is a measurement bin determined by the finite time-frequency register. Since the readout distribution consists of probabilities, we work with the squared kernel
\begin{equation}
    \label{eq:kernel}
    K(\theta) = |D_M(\theta)|^2 = \left|\frac{\sin(\pi M \theta)}{M \sin(\pi \theta)} \right|^2, \quad \theta \in [-\frac{1}{2}, \frac{1}{2}]
\end{equation}
where the Dirichlet kernel $D_M$ arises naturally from the discrete Fourier transform over $M=2^m$ time steps used in the DOS-QPE protocol. Note that the dividing factor $M$ normalizes the kernel integral to 1 in the large-$M$ limit, ensuring the convergence of $D_M$ to a Dirac delta in the same limit. The final observed probability distribution that we aim to reconstruct is
\begin{equation}
    \label{eq:distro}
    \tilde{P}(\theta) = \sum_{i=1}^{R}d_i \cdot K(\theta - \theta_i) .
\end{equation}


\subsection{Compressed sensing and optimization}
\label{sec:optimization}
To recover the discrete phases $\{\theta_i\}$ and their degeneracies $\{d_i\}$ from the noisy coarse-grained histogram $\{P_i\}$, we reframe the degeneracy integer problem as an inverse problem. The goal is to infer a sparse spectral signal --- convolved with a known kernel --- from its noisy, finite-resolution observation. This setting naturally invites tools from convex optimization to minimize the discrepancy between the observed histogram and a model spectrum.

We formulate the problem on a grid of candidate phases $\{\theta'_j\} \subset [0,1)$, finer than the hardware-limited resolution $2^{-m}$. We obtain an underdetermined system where the number of potential spectral components exceeds the number of measured bins. To select a physically meaningful sparse solution, we employ a quadratic program~\cite{Boyd_2004} with an $\ell_1$ regularization penalty (basis-pursuit, LASSO)~\cite{Hastie_2015}, which encourages sparsity by penalizing the absolute sum of the degeneracy coefficients. This results in a practical realization of compressed sensing~\cite{Donoho_2006, Cand_s_2006}, a framework for recovering sparse signals from limited or noisy measurements. The finer grid provides an overcomplete dictionary of possible phases, while the $\ell_1$ penalty enforces that only a small subset is selected. The resulting optimization routine enables a super-sampled spectral reconstruction despite a coarser hardware discretization~\cite{Bertsimas_2024}.

We introduce a uniform grid of $G$ candidate phases $\{\theta'_i\}_{i=1}^G$, producing the dictionary matrix $A\in\mathbb{R}^{N\times G}$:
\begin{equation}
A_{i,j} = K(\theta_i - \theta'_j).
\end{equation}
Writing $\mathbf{P}=(P_k)_{k=1}^N$ and $\mathbf{w}=(w_j)_{j=1}^G$ as the degeneracy coefficients for the grid points, the forward model is
\begin{equation}
\mathbf{P} \approx A\mathbf{w}.
\end{equation}
The problem naturally maps onto a mixed-integer quadratic program~\cite{Nesterov_2004, Conforti_2014} with convex, non-linear objective function $L_{2, \text{MI}} = \min_{\mathbf{w}\in\mathbb{N}_+}|A\mathbf{w}-\mathbf{P}|_2^2$ with $\sum_{j=1}^G w_j = 2^n$. Quadratic programming is NP-complete~\cite{Vavasis_1990} in general, with the variant of mixed-integer programming (MIP) enforcing $w_j\in\mathbb{Z}+$ and $\sum_jw_j=\sum_i d_i=2^n$ being NP-hard~\cite{Pia_2016, Renegar_1992, Liberti_2019}. For this reason, we relax the problem from integer degeneracies to real and add a further final step of clustering of phases and integer approximation on the solution. This makes the problem a quadratic program with a convex objective function, which, together with linear programs, is in P~\cite{Kozlov_1980, Renegar_1992, Liberti_2019}. The final objective function we optimize against is
\begin{equation}
    \label{eq:L2}
    L_2 = \min_{\mathbf{w}\in\mathbb{R}_+^G}|A\mathbf{w}-\mathbf{P}|_2^2 + \lambda|\mathbf{w}|_1 \quad \text{s.t.}\quad \sum_{j=1}^G w_j = 2^n ,
\end{equation}
where we added the $\ell_1$ regularization penalty controlled by $\lambda>0$. The squared-$L_2$ term ensures data fidelity, and the $\ell_1$ term promotes sparse support in the candidate-phase weights. This convex QP can be solved efficiently at moderate $G$ using off-the-shelf solvers. In our simulations, we do this using the CLARABEL solver~\cite{clarabel} and CVXPY~\cite{cvxpy, cvxpy2}.

Note that, because the quadratic part of $L_2$ is piecewise linearizable, we can also reformulate the problem as a linear program $L_1 = \min_{\mathbf{w},\mathbf{r}}(\mathbf{r}+\lambda\mathbf{w})$ such that $|A\mathbf{w}-\mathbf{P}| < \mathbf{r}$ with $\mathbf{w}\in\mathbb{R}_+^G$ and $\sum_{j=1}^G w_j = 2^n$.

The solution $\mathbf{w}^*$ of Eq.~\eqref{eq:L2} is typically sparse in the Hilbert space of dimension $2^n$. Consider the vector $\bm{J}=j : w_j^* > \tau$ of those phase indices whose solution degeneracy is above a certain target threshold $\tau$. To recover the final phases and integer degeneracies, we apply the following steps:
\begin{enumerate}
    \item Thresholding. We retain grid indices in $\bm{J}$ and their weights $w_j^*$.
    \item Clustering. We merge nearby phases $\phi_j$ within a tolerance $\epsilon$ via single-linkage, forming clusters $C_1, \ldots, C_{\max\bm{J}}$. This $\epsilon$ can be taken of the order of the initial grid size of DOS-QPE defined by the time-frequency register dimension.
    \item Rounding. For each cluster $C$, we compute
    \begin{equation}
        \hat\theta = \frac{\sum_{j\in C} w_j^\star\psi_j}{\sum_{j\in C} w_j^\star}, \quad \hat d = \left\lfloor\sum_{j\in C} w_j^\star\right\rceil .
    \end{equation}
\end{enumerate}
In this way, we restore integer degeneracies and obtain a final estimate $\{\hat\theta_i\}_{i=1}^{\max\bm{J}}$ with corresponding $\{\hat d_i\}_{i=1}^{\max\bm{J}}$.

To quantify the reconstruction error, we use the 1-Wasserstein distance~\cite{Kantorovich_1960} $W_1(\mu, \nu)$ between discrete spectral distributions $\mu = \sum_i d_i \, \delta_{\theta_i}$ and $\nu = \sum_j \hat d_j \, \delta_{\hat\theta_j}$, where $\delta_\theta$ denotes the Dirac measure at $\theta$. This metric is defined as
\begin{equation}
    W_1(\mu, \nu) = \inf_{\gamma \in \Gamma(\mu, \nu)} \sum_{i,j} \gamma_{ij} \, |\theta_i - \hat\theta_j|,
\end{equation}
where $\Gamma(\mu, \nu)$ is the set of all couplings between $\mu$ and $\nu$ that preserve total weight. Intuitively, this metric quantifies the dissimilarity between the two distributions by measuring the minimum amount of ``mass'' times ``travel'' that needs to be moved to shape one into the other~\cite{Villani_2009}. In the following, we compute $W_1$ between the sampled spectrum and the exact one, as well as between the optimized solution and the exact spectrum. This provides a principled way to evaluate the fidelity of spectral estimation, taking into account both eigenphase positions and degeneracies.


\section{Applications}
\label{sec:applications}

In the following, we show three applications of DOS-QPE: the Fermi-Hubbard model, an electronic structure Hamiltonian, and a nuclear Hamiltonian. We use $\text{t}|\text{ket}\rangle$~\cite{sivarajah2020t} to construct the circuits, Qulacs for state vector evaluation~\cite{suzuki2021qulacs}, and OpenFermion for constructing qubit Hamiltonians~\cite{mcclean2020openfermion}. To evaluate the molecular integrals in the electronic structure section, we use the PySCF Python package~\cite{sun2018pyscf}.


\subsection{The Fermi-Hubbard Model}
\label{sec:hubbardcase}

The Fermi-Hubbard model~\cite{hubbardI,hubbardII,hubbardIII,hubbardIIII} is one of the standard models of condensed matter physics as it describes relevant features of strongly-correlated electronic systems. It has also become a standard benchmark for Hamiltonian simulation algorithms~\cite{kan2024resource,babbush2018encoding,arute2020observation} in the quantum computing community since exact solutions can be found for the one-dimensional case~\cite{lieb1968absence}. The study of the two- and higher-dimensional cases is a topic of intense research activity, where only numerical simulations are available~\cite{chen2021quantum,varney2009quantum,pasqualetti2024equation}. Its Hamiltonian is written as
\begin{equation}
    H = -t \sum_{\langle i, j\rangle, \sigma} a^\dagger_{i\sigma} a_{j\sigma} + U\sum_i n_{i\uparrow} n_{i\downarrow},
\end{equation}
where the quantity $\sigma=\uparrow,\downarrow$ denotes spin projection, $n_{i\sigma} = a_{i\sigma}^\dagger a_{i\sigma}$ is a particle-number operator, the symbol $\langle i, j \rangle$ restricts the sum over the indices of neighboring lattice sites and $t$ and $U$ are the hopping and repulsion parameters, respectively. 

In the remainder of this section, we apply the DOS-QPE algorithm to the Fermi-Hubbard chain model with three sites. We choose $t=1$ and $U=4$ such that we are in the so-called ``intermediate coupling'' regime, under which it is difficult classically to extract even ground-state properties from numerical simulations at half-filling~\cite{fradkin2015colloquium}. We can then map this second-quantized Hamiltonian, using the Jordan-Wigner encoding, to a linear combination of products $P$ of Pauli operators $\sigma_j$,
\begin{equation}
    {H} = \sum_i h_i {P}_i = \sum_i h_i \bigotimes_j {\sigma}^i_j,
\end{equation}
where $h_i$ are the coefficients in the transformation.

\begin{figure}
    \centering
    \includegraphics[width=\linewidth]{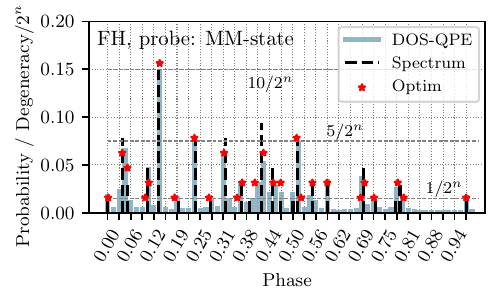}
    \caption{\textbf{DOS-QPE on the Fermi-Hubbard model.} We run the DOS-QPE circuit on a Fermi-Hubbard toy model with three sites, $t=1$, and $U=4$, employing $m=6$ ancillas to resolve the spectrum. In the time-frequency register, we obtain the probability distribution of the DOS (blue bars). We compare this with the exact spectrum (black bars) of the Hamiltonian. While the dimensionality of the time-frequency register is insufficient to resolve the spectral gaps of the spectrum, we show that quadratic programming aided by compressed sensing helps in the phase and degeneracy reconstruction (red stars). The 1-Wasserstein distance of the optimized solution improved by $\sim51\%$ over the integer-rounded sampled distribution.}
    \label{fig:fermi_hubbard}
\end{figure}

We run the DOS-QPE circuit of Fig.~\ref{fig:dosqpe}a) and show the results in Fig.~\ref{fig:fermi_hubbard}. For this simulation, we have state and purification registers of dimension $n=6$, and we try to resolve the spectrum with a time-frequency register of dimension $m=6$, for a total of 18 qubits. The Hamiltonian is rescaled following App.~\ref{sec:rescaling_appendix} and the unitary propagation is done via fourth-order Trotterization. Comparing the DOS $S(\theta)$ of the Hamiltonian $H$ to its exact spectrum, we note that the dimensionality of the time-frequency register is insufficient for a perfect spectrum reconstruction. That is why we employ the quadratic program and compressed sensing technique explained in Sec.~\ref{sec:optimization}. Using such convex optimization, we can obtain a better estimate of the integer degeneracies and the real phases of the spectrum. As described in Sec.~\ref{sec:optimization}, we use the 1-Wasserstein distance against the exact spectrum to quantify the deviation of the distributions, finding $\sim16\%$ improvement over the sampled distribution and more than $\sim51\%$ over the corresponding integer-rounded version. 

\subsection{Electronic Structure}
\label{sec:molecularcase}
The application of quantum algorithms to chemistry typically begins with the construction of the system Hamiltonian, written in the second-quantization formalism as,
\begin{equation}\label{eq:sq_hamiltonian}
    {H} = \sum_{pq} h_{pq} a_p^\dagger a_q + \frac12 \sum_{pqrs} h_{pqrs}a_p^\dagger a_q^\dagger a_s a_r.
\end{equation}
The constants $h_{pq}$ and $h_{pqrs}$ are the integrals,
\begin{equation}
    h_{pq} = \int \text{d}\mathbf{x}\phi_p^*(\mathbf{r}) \left( -\frac{\nabla^2}{2} - \sum_A \frac{Z_A}{|\mathbf{r} - \mathbf{R_A}|} \right) \phi_q(\mathbf{r}) 
\end{equation}
and 
\begin{equation}
    h_{pqrs} = \int \text{d}\mathbf{r}_1  \text{d}\mathbf{r}_2 \frac{\phi_p^*(\mathbf{r}_1) \phi_q^*(\mathbf{r}_2) \phi_r(\mathbf{r}_2) \phi_s(\mathbf{r}_1)}{|\mathbf{r}_1 - \mathbf{r}_2|},
\end{equation}
respectively, and are routinely obtained as part of a classically inexpensive Hartree-Fock calculation. The quantities $\phi_p$ are the molecular orbitals and form the 1-particle wavefunctions mentioned in Sec.~\ref{sec:dicke_states}, $\mathbf{r}$ are the positions of the electrons, and $Z_A$ and $\mathbf{R_A}$ are the charges and positions of the nuclei, respectively. For the same reasons as in Sec.~\ref{sec:dicke_states} and the former section, the Jordan-Wigner mapping is selected.

In most classical computational chemistry studies, the number of electrons in a system remains fixed. Descriptions of many chemical phenomena are framed in terms of energy differences as the nuclear configuration changes, or between eigenstates at some fixed geometry. Even in cases where the number of electrons varies, the full spectrum for all eigenvalues of all occupation numbers is rarely required. As such, the Dicke state $|D^n_k\rangle$, where $k$ is set to the number(s) of electrons in the system, is the most relevant probe for the DOS-QPE algorithm as it is applied in this section.

\begin{figure*}
    \centering
    \includegraphics[width=\linewidth]{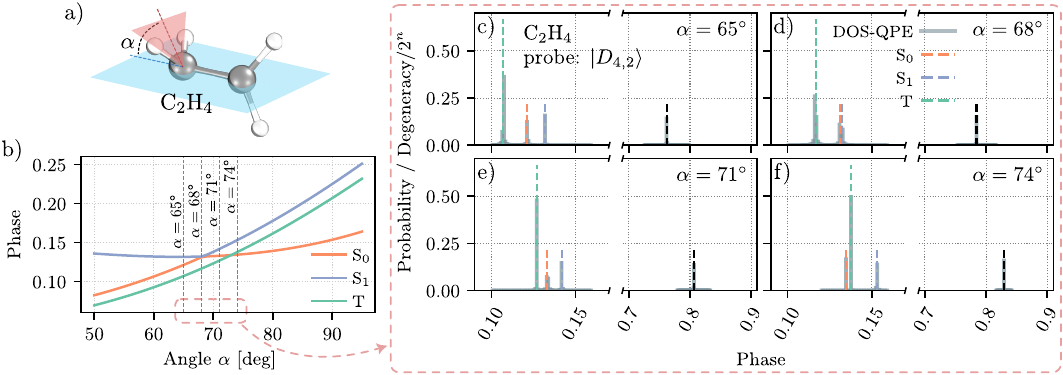}
    \caption{\textbf{DOS-QPE on the electronic structure of ethylene.} a) The structure of the ethylene molecule showing a twisted conformation and the angle $\alpha$ which is varied during the scan of the potential energy surfaces. b) Plot of the lowest three eigenvalues of the Hamiltonian, scaled such that the eigenvalues lie in the interval $[0, 1)$. The green line corresponds to the triplet state $T$, the orange line to the lowest lying singlet state S$_0$ and the purple line to the first excited singlet state S$_1$. c--f) The distributions obtained with DOS-QPE with $\alpha= 65^\circ, 68^\circ, 71^\circ$ and $74^\circ$, respectively. In each plot c--f), the colour of each vertical dashed line corresponds to the phases obtained via exact diagonalization and matches those used in b). The height of the dashed lines shows the height the peaks in the distribution would be if all eigenvalues were to be exact integer multiples of $\frac{1}{N_\text{anc}}$.}
    \label{fig:ethylene}
\end{figure*}

We choose the twisted conformation of ethylene as the example chemical system, since by varying the pyramidalization angle we expect to encounter a conical intersection between the two lowest lying singlet eigenstates, and a crossing of the lowest lying singlet state with the triplet ground state~\cite{xu2025conical}. We first optimise the planar geometry at the CCSD(T) level classically in the cc-pVDZ basis, then rotate the two hydrogens on one carbon atom to have dihedral angles with the hydrogen atoms on the other carbon atom of exactly $90^\circ$. We then vary the angle $\alpha$ as shown in Fig.~\ref{fig:ethylene}a). We select a 2-electrons in 2-orbitals active space, and along the reaction coordinate, evaluate the overlap of the orbitals at each point with the orbitals from the previous point, such that we can place the orbitals at maximum coincidence at each step along the potential energy surfaces and ensure a consistent orbital selection. Such that the eigenvalues lay in a fixed interval and the movement of each peak in Fig.~\ref{fig:ethylene}c--f)  is smooth, and to obtain a smooth curve in Fig.~\ref{fig:ethylene}b), we find our target Hamiltonians with the following formula,

\begin{equation}
    H_{\text{target}} = (1-\delta) \left(\frac{H - \lambda_{\text{min},  50^\circ} + \delta}{\lambda_{\text{max},50^\circ} -\lambda_{\text{min},  50^\circ}}\right),
\end{equation}
where in all cases $\delta=0.05$, which serves as a shift to move the lowest phase away from zero and simplify communication regarding Fig.~\ref{fig:ethylene}, and  $\lambda_{\text{min},  50^\circ}
$ and $\lambda_{\text{max},50^\circ}$ are the lowest and highest lying 2-particle eigenvalues obtained by exact diagonalization of the geometry at $\alpha = 50^\circ$, respectively.

The results for exact diagonalization of the active space Hamiltonian are shown in Fig.~\ref{fig:ethylene}b). We observe that the conical intersection is at approximately $68^\circ$. We then select three evenly spaced points around the approximate location of the conical intersection, at $65^\circ, 71^\circ$, and $74^\circ$, at which we perform state-vector DOS-QPE calculations. We use second-order Trotterization with one time step for the preparation of $U_H$ and, as previously mentioned, set $U_\psi$ to be the $|D^4_2\rangle$ state-preparation unitary. We also marginalise the exact probabilities of measurement outcomes on the state and purification qubit registers such that the results we obtain correspond to the exact probabilities of the measurement outcomes on the ten ancilla qubits.

By restricting the probe to being an even superposition of only $^4C_2$ states, we expect to observe a maximum of $6$ distinct eigenvalues. Spin symmetry simplifies the picture since we know that for any triplet state with total spin $S=1$, there are three degenerate states with $M_s=-1, 0, 1$. We can infer the multiplicity of the states corresponding to the phases, then, from the relative peak heights in the spectra shown in Fig.~\ref{fig:ethylene}c--f): each contains three degenerate triplet eigenstates with different $M_s$ values, and three peaks corresponding to singlet states which have no degeneracy by spin symmetry, totaling six states as expected. In what remains of this section, we will refer to the lowest and second-lowest singlet state as S$_0$ and S$_1$, respectively, and to the triplet state as T. The other singlet state (marked with a black dashed line) is not of any interest in this discussion but is included in the figure for completeness.

We see by comparing the phases shown in Fig.~\ref{fig:ethylene}b) and the peaks in Fig.~\ref{fig:ethylene}c) that we gain access to the correct ordering of states, the phases agreeing with exact diagonalization. Increasing $\alpha$ to $68^\circ$ (see Fig.~\ref{fig:ethylene}d)), close to the conical intersection between S$_0$ and S$_1$, we see near degeneracy between the phases corresponding to the S$_0$ and S$_1$ states, as expected, and an increase in the phase found for the $T$ state. Further increasing $\alpha$ to $71^\circ$, shows the separation between the phases corresponding to S$_0$ and S$_1$ return, as is the case in Fig.~\ref{fig:ethylene}b). The distribution shown in Fig.~\ref{fig:ethylene}f) shows that the phases corresponding to T and S$_0$ have crossed, and the ground state now carries singlet multiplicity, as expected from the results of the exact diagonalization.

From the faithful reconstruction of the spectra along the potential energy surfaces, we conclude that DOS-QPE represents a valuable tool for the fault-tolerant era of quantum computation for the study of chemical systems. In the form it is presented here, we anticipate DOS-QPE being particularly useful for investigating phenomena arising directly from more than one eigenstate at any given molecular geometry. An interesting next step for this approach is to leverage other known physical symmetries to construct more targeted probe states.


\subsection{Nuclear Structure}
\label{sec:nuclearcase}

The description of nuclear structure in atomic nuclei follows a similar prescription as that of molecules in quantum chemistry within the nuclear shell model~\cite{caurier2005shell}. Under this approach, protons and neutrons (collectively called \emph{nucleons}), basic fermion components of a nucleus, move in a set of single-particle states that form a valence space. These single-particle states, also called orbitals, are labeled by the quantum numbers $(n,l,j,m,t_z)$, where $n$ is the principal quantum number, $l$ is the orbital angular momentum, $j$ is the total angular momentum after coupling with the nucleon spin$-\frac{1}{2}$, $m$ is its third-component projection and $t_z$ is the third-component projection of the \emph{isospin} $t = \frac{1}{2}$, an additional quantity equivalent to the spin defined in order to discern between protons and neutrons. 

The nuclear shell model can be described by means of a second-quantized Hamiltonian as in Eq.~\eqref{eq:sq_hamiltonian}, where the single-particle energies $h_{pp}$ and two-body matrix elements $h_{pqrs}$ are adjusted to reproduce observable properties of selected nuclei~\cite{poves1981theoretical}. The nuclear states are described by a well-defined total angular momentum $J$ and isospin $T$, resulting from the coupling of individual nucleonic $j$ and $t$; with their corresponding third-component projections $M$ and $T_z$, which are equal to the sum of the third components of the nucleons in the valence space, $m$ and $t_z$. Thus, a many-body basis state can be written in terms of states with well-defined $M$ and $T_z$ 
\begin{equation}
    \ket{JMTT_z} = \sum_{i} c_{i}\ket{i, MT_z},
\end{equation}
where $i$ labels additional quantum numbers needed for the description of the many-body state, and $c_i$ are the coefficients in the superposition. The Hamiltonian matrix is then built using this many-body basis, and the nuclear states can be found as its eigenstates. Because of the two kinds of particles in a nucleus, protons and neutrons, the scaling of this many-body basis is as $\binom{ d_\text{sh} }{ N_\text{val} } \binom{ d_\text{sh} }{ Z_\text{val} }$, where $d_\text{sh}$ is the dimension of the valence space and $N_\text{val}$, $Z_\text{val}$ are the valence neutron and proton numbers, respectively. 

Many quantum algorithms have been applied to obtain nuclear-structure properties based on the ground state~\cite{kiss2022quantum,dumitrescu2018cloud,romero2022solving,perez2023nuclear,stetcu2022variational,costa2024quantum,sarma2023prediction,carrasco2025comparison} and also low-lying excited states~\cite{zhang2025excited,bhoy2024shell,ruiz2022accessing,robin2023quantum}. The structure and entanglement features of nuclear states have also been thoroughly studied~\cite{robin2021entanglement,hengstenberg2023multi,perez2023quantum,tichai2023combining}.   

\begin{figure}
    \centering
    \includegraphics[width=\linewidth]{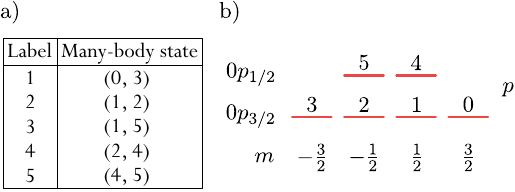}
    \caption{\textbf{a) Many-body state basis used for ${}^6$Be.} Label represents the indices in the corresponding Hamiltonian matrix in this many-body basis. The many-body states are composed of single-particle states from all possible combinations of two valence protons in the $p-$shell variational space of panel b) satisfying the condition $M=m_1+m_2=0$. \textbf{b) Configurational valence space for the $p-$shell.} The valence space is decomposed in two degenerate $j$ orbitals with degeneracy $(2j+1)$. The labels on the left correspond to the standard spectroscopic notation $n\ell_j$, where $\ell=p$ means $\ell=1$. Numbers on top of every single-particle state represent the labels used in this work. Slightly modified from~\cite{perez2023nuclear}.}
    \label{fig:pshell}
\end{figure}

In this work, we aim to resolve the full spectrum in a single calculation with DOS-QPE. We simulate the ${}^6$Be nucleus, which is composed of two protons in the valence space of a \emph{p} shell, as shown in Fig.~\ref{fig:pshell}, with six single-particle states for the protons. Thus, we have in this case two identical particles moving in the valence space, and consequently, the total isospin is $T=1$. The Hamiltonian used is based on the phenomenological Cohen-Kurath interaction~\cite{Cohen_Kurath}.

While we could directly transform the second-quantized Hamiltonian using the Jordan-Wigner mapping as demonstrated in the former examples, for this specific case we opt to build the Hamiltonian matrix using the minimal many-body basis spanned for ${}^6$Be, consisting of only five states whose total angular momentum projection $M$ is equal to zero~\cite{perez2023nuclear}. The many-body basis used for ${}^6$Be is listed in the table of Fig.~\ref{fig:pshell}a). The resulting Hamiltonian matrix is then mapped onto the qubit operator basis using the Linear Combination of Unitaries (LCU) formalism~\cite{childs2012hamiltonian}. Thus, using this mapping, we managed to reduce the size of the Hamiltonian from six to three qubits, at the expense of three additional degenerate eigenvalues equal to zero.

\begin{figure}
    \centering
    \includegraphics[width=\linewidth]{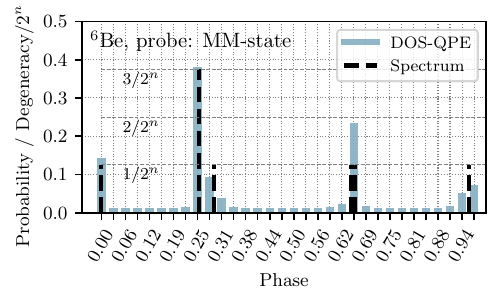}
    \caption{\textbf{DOS-QPE on $^6$Be.} The result of DOS-QPE with a maximally mixed state probe. Sixth-order Trotterization of the unitary evolution and five qubits in the time-frequency register were used for these results. We observe that, while the distribution (blue bars) faithfully captures the exact eigenspectrum, it is not fine-grained enough to resolve the almost-degenerate peaks at phase $\sim 0.65$, and more ancillary qubits would be needed. The tallest peak represents the redundant three-times degenerate zero eigenvalue resulting from building the nuclear Hamiltonian matrix using the LCU approach.}
    \label{fig:6be-dosqpe}
\end{figure}

The results of the DOS-QPE calculation for this system are shown in Fig.~\ref{fig:6be-dosqpe}. The calculation was performed using a sixth-order Trotterization of the unitary propagation and five ancillary qubits in the time-frequency register for the spectrum resolution. We observe that, under these resources, four out of five physical phases (eigenvalues) of $^6$Be are retrieved within a precision dictated by the number of ancillary qubits used (see App.~\ref{sec:dosqpe_appendix}). The resolution for the almost degenerate peaks at $\sim 0.65$ is not enough and the number of ancillary qubits in the time-frequency register needs to be increased. A spurious zero eigenvalue with degeneracy of three is also found as an LCU artifact used to map the Hamiltonian to a qubit-operator basis.

In light of these results, although demonstrated on a small example, we may conclude that DOS-QPE could become a promising tool for efficiently extracting spectral information and thus dynamical processes~\cite{roggero2020preparation,stetcu2025quantum,garcia2024nuclear,zhang2025quantum,du2025systematic} in atomic nuclei for near-term quantum devices.


\section{Discussion and conclusion}
We have shown how, by using a generalization of quantum phase estimation probed with maximally mixed or symmetry-adapted ensemble states, we can access spectral information of the generating Hamiltonians. By embedding QPE within a purified framework~\cite{scali2025signed}, the traditional eigenstate-dependent algorithm transforms into a flexible tool for ensemble-based spectral analysis. This approach grants access to the density of states of Hamiltonians, effectively revealing their spectral properties. From here, thermodynamic and dynamical properties of systems become accessible while mitigating the effects of noisy or imperfect state preparations.

We framed the recovery of discrete spectral features from the output of DOS-QPE as a convex optimization problem. Exploiting sparsity and physical constraints, we resolve spectral structures and obtain high-fidelity reconstructions, and we propose a modular postprocessing pipeline that restores physically meaningful integer-valued spectra. Together with the circuit primitive, this yields an end-to-end workflow from mixed-state probes to reconstructed eigenspectra.

While we have demonstrated DOS-QPE as a flexible primitive for ensemble-based quantum spectroscopy across fermionic and nuclear models, further optimizations and resource reductions are desirable. One immediate goal is to reduce circuit complexity and resource cost, tailoring the primitive for currently available quantum hardware~\cite{fujitsu_qpu}. Implementing single-ancilla variants~\cite{Ding_2023,Ding_2024_2,Ding_2024_3} or taking advantage of shadow techniques~\cite{Huang_2020,Elben_2022,Faehrmann_2025} could significantly lower circuit depth and overhead, making the primitive viable for near-term applications.

Additional directions include exploring alternative probe states, such as spin-selective or other symmetry-constrained ensembles, which may target subspace-resolved spectral features often inaccessible to conventional methods. In this context, developing spectrum manipulation techniques to enhance signal contrast, reduce costs, and mitigate noise could be particularly valuable, accelerating the reconstruction process and alleviating current bottlenecks in measurement and post-processing.

DOS-QPE thus emerges as a framework that bypasses eigenstate-preparation bottlenecks while leveraging the incoherent nature of realistic environments to return physically relevant observables that are otherwise difficult to access. Conceptually, this purified viewpoint resonates with the random purification channel paradigm developed in quantum learning theory by Tang et al.~\cite{tang}, Pelecanos et al.~\cite{pelecanos}, and Girardi et al.~\cite{girardi}: in both cases, purification elevates mixed-state access from a nuisance to a resource for running pure-state algorithms. Connections to non-linear functions of spectral densities, gradient estimation, and Bell-type measurements~\cite{Scali_2024_2} indicate a broader algorithmic utility within hybrid optimization and quantum learning frameworks. Embedding DOS-QPE within Green's function formulations~\cite{Libbi_2022,Umeano_2025} could further support dynamical response theory, offering access to electronic, vibrational, or nuclear observables and strengthening the role of purified phase estimation as a core primitive for future quantum simulations.


\section*{Author contributions}
S.S. and J.K. had the idea and developed the methods of the paper. J.K. wrote Dicke State, FH and Electronic Structure sections. A.M.R. developed the nuclear application and wrote the relative section. S.S. linked the paper to the Random Purification Channel, wrote the rest of the manuscript including appendices and prepared all figures. All authors reviewed the manuscript.

\section*{Funding}
This work has been entirely funded by Fujitsu Research of Europe.


\appendix


\section{QPE - derivation}
\label{sec:qpe_appendix}

Given a unitary operator $U$ and an eigenvector $|\psi\rangle$ such that
\begin{equation}
    U|\psi\rangle = e^{2\pi i \phi}|\psi\rangle, \quad \phi \in [0,1),
\end{equation}
the goal is to estimate $\phi$ to $m$-bit precision. The registers we find in a standard implementation of QPE are
\begin{itemize}
    \item Time-frequency register: $m$ qubits, initialized to $|0\rangle^{\otimes m}$.
    \item State register: $n$ qubits, initialized to a target eigenvector $|\psi\rangle$.
\end{itemize}

The entire initial state facing the circuit is,
\begin{equation}
    |\Psi_0\rangle = |0\rangle^{\otimes m} \otimes |\psi\rangle.
\end{equation}
The first operation of QPE is the application of a layer of Hadamard gates $H^{\otimes m}$ to the first register to create the uniform superposition,
\begin{equation}
    |\Psi_1\rangle = \left( \frac{1}{\sqrt{2^m}} \sum_{k=0}^{2^m-1} |k\rangle \right) \otimes |\psi\rangle.
\end{equation}
We then apply the unitary operation $U_\text{QPE}$, consisting of a controlled-$U^{2^j}$ operation for each qubit $j$ in the time-frequency register, $|\Psi_2\rangle = 2^{-m/2} \sum_{k=0}^{2^m-1} |k\rangle \otimes U^k|\psi\rangle $.
Since $U^k|\psi\rangle = e^{2\pi i k\phi}|\psi\rangle$, this results in
\begin{equation}
    |\Psi_2\rangle = \frac{1}{\sqrt{2^m}} \sum_{k=0}^{2^m-1} e^{2\pi i k\phi} |k\rangle \otimes |\psi\rangle.
\end{equation}
We then apply the inverse quantum Fourier transform $U_\text{QFT}^\dagger$ to the time-frequency register. The Fourier basis state $2^{-m/2} \sum_{k=0}^{2^m-1} e^{2\pi i k\phi} |k\rangle$
is mapped to the computational basis state $|\tilde{\phi}\rangle$, where $\tilde{\phi}$ is the best $m$-bit approximation of $\phi$. If $\phi$ has an exact $m$-bit binary expansion $\phi = 0.\phi_1\phi_2\ldots\phi_m$, then
\begin{equation}
    U_\text{QFT}^\dagger\left( \frac{1}{\sqrt{2^m}} \sum_{k=0}^{2^m-1} e^{2\pi i k\phi} |k\rangle \right) = |\phi_1\phi_2\ldots\phi_m\rangle.
\end{equation}
The final state of the system is $|\Psi_{\text{final}}\rangle = |\tilde{\phi}\rangle \otimes |\psi\rangle$ and measuring the time-frequency register gives $\tilde{\phi}$, the estimate of $\phi$.

\subsection*{Error}

Note that, when $\phi$ cannot be exactly represented with $m$ bits, we have
\begin{equation}
    \phi = \frac{s}{2^m} + \delta, \quad \text{with} \quad |\delta| < \frac{1}{2^{m+1}}.
\end{equation}
The probability of measuring $s$, the nearest $m$-bit approximation to $\phi$, is
\begin{equation}
    P_{\text{success}} \geq \left( \frac{\sin(\pi \delta 2^m)}{2^m \sin(\pi \delta)} \right)^2,
\end{equation}
that is exactly 1 if $\delta = 0$, while, for small $\delta$, the phase estimation error is bounded by
\begin{equation}
    |\tilde{\phi} - \phi| \leq \frac{1}{2^{m+1}} \quad \text{(with high probability)}.
\end{equation}


\section{DOS-QPE - derivation}
\label{sec:dosqpe_appendix}

Consider the case where the input to QPE is not an eigenstate of $U$, but rather an arbitrary mixed state
\begin{equation}
    \rho_S = \sum_{j=1}^N p_j \,|\psi_j\rangle\langle \psi_j|,
\end{equation}
with $\{|\psi_j\rangle\}$ the eigenbasis of $U$ satisfying $U|\psi_j\rangle = e^{2\pi i \phi_j}|\psi_j\rangle$.  We purify $\rho_S$ by introducing an auxiliary register of the same dimension. The purification is
\begin{equation}
    |\Psi\rangle_{SP} = \sum_{j=1}^N \sqrt{p_j}\;|\psi_j\rangle_S \otimes |j\rangle_P,
\end{equation}
where $\{|j\rangle\}_P$ is an orthonormal basis of the purification register. For simplicity, we drop the subscripts in the following. The total system includes three registers:
\begin{itemize}
    \item Time-frequency register: $m$ qubits, initialized to $|0\rangle^{\otimes m}$.
    \item State register: $n$ qubits, initialized to the system of interest.
    \item Purification register: $n$ qubits, entangled with the state register to create the purification.  This register is to be traced out at the end.
\end{itemize}

The initial state is
\begin{equation}
    |\Psi_0\rangle
    = |0\rangle^{\otimes m} \otimes \left(\sum_{j=1}^N \sqrt{p_j}\;|\psi_j\rangle \otimes |j\rangle\right).
\end{equation}
Applying Hadamard gates $H^{\otimes m}$ to the time-frequency register creates the superposition
\begin{equation}
    |\Psi_1\rangle
    = \left(\frac{1}{\sqrt{2^m}} \sum_{k=0}^{2^m-1} |k\rangle\right)
      \otimes \left(\sum_{j=1}^N \sqrt{p_j}\;|\psi_j\rangle \otimes |j\rangle\right).
\end{equation}
We then apply the controlled-unitary operation $U_{\text{QPE}}$, where each $|k\rangle$ controls $U^k$ on the state register.  Since $U^k|\psi_j\rangle = e^{2\pi i k \phi_j}|\psi_j\rangle$, the resulting state is
\begin{equation}
    |\Psi_2\rangle
    = \frac{1}{\sqrt{2^m}} \sum_{k=0}^{2^m-1} |k\rangle
      \otimes \sum_{j=1}^N \sqrt{p_j}\,e^{2\pi i k \phi_j}\;|\psi_j\rangle \otimes |j\rangle.
\end{equation}
Next, we apply the inverse quantum Fourier transform $U_{\text{QFT}}^\dagger$ to the time-frequency register.  The Fourier transform maps
\begin{equation}
    \frac{1}{\sqrt{2^m}} \sum_{k=0}^{2^m-1} e^{2\pi i k \phi_j} |k\rangle
    \;\mapsto\;
    \sum_{y=0}^{2^m-1} c_j(y)\;|y\rangle,
\end{equation}
where
\begin{equation}
    c_j(y) = \frac{1}{2^m} \sum_{k=0}^{2^m-1}
    e^{2\pi i k \bigl(\phi_j - \tfrac{y}{2^m}\bigr)},
\end{equation}
where each outcome $y\in\{0, \ldots, 2^{m}-1\}$ corresponds to a phase estimate $\tilde{\phi_j} = y/2^m$. After the inverse Fourier transform, the state becomes
\begin{equation}
    |\Psi_3\rangle
    = \sum_{j=1}^N \sqrt{p_j}\;\sum_{y=0}^{2^m-1}
    c_j(y)\;|y\rangle \otimes |\psi_j\rangle \otimes |j\rangle.
\end{equation}
By tracing out the purification register, the measurement of the time-frequency register gives the outcome $y$ with probability
\begin{equation}
    P(y) = \sum_{j=1}^N p_j\,\bigl|c_j(y)\bigr|^2,
\end{equation}
which corresponds to sampling from the spectral measure of $\rho_S$ under $U$, coarse-grained to resolution $\Delta\phi = 1/2^m$. Each measurement outcome $y$ yields a phase estimate $\tilde{\phi_j} = y/2^m$, and the empirical distribution of $\tilde{\phi}_j$ reconstructs the density of states of $U$.

\subsection*{Error and sampling}
\label{sec:dosqpe_error_appendix}

Three main sources of error affect this procedure:
\paragraph{Resolution error} The Fourier truncation limits the phase resolution to
\begin{equation}
    \Delta\phi = \frac{1}{2^m}.
\end{equation}

\paragraph{Statistical sampling error} Each DOS-QPE run yields one sample from the spectral distribution. After $M$ shots, the per-bin standard deviation is defined by the binomial sampling error bound, that is
\begin{equation}
    \epsilon \approx \sqrt{ \frac{p(1-p)}{M}} \leq \frac{1}{2\sqrt{M}},
\end{equation}
assuming worst-case $p = 0.5$. The total $L_2$ error in the reconstructed histogram $\hat{\rho}_S$ is
\begin{equation}
    \|\hat{\rho}_S - \rho_S\|_2 \sim \sqrt{ \frac{B}{M} } = \sqrt{ \frac{2^m}{M} },
\end{equation}
where $\hat{\rho}$ is the empirical distribution (histogram) extracted and $B = 2^m$ is the number of bins in such distribution. To achieve a total error $\delta$, the number of shots must satisfy
\begin{equation}
    M \gtrsim \frac{2^m}{\delta^2}.
\end{equation}

\paragraph{Scaling with Hilbert space dimension}

Let $N_{\text{eff}}$ denote the number of eigenstates contributing non-trivially to $\rho_S$:
\begin{itemize}
    \item For a \emph{maximally mixed state}, $\rho_S = \mathbb{I}/2^n$, so $N_{\text{eff}} = 2^n$, and the DOS-QPE samples the full spectrum of $U$.
    \item For a \emph{restricted ensemble}, such as the uniform mixture of Dicke states, the support of $\rho_S$ is reduced. For example, fixed Hamming-weight Dicke states span a subspace of dimension $\binom{n}{k} \sim \mathcal{O}(n^k)$.
    \item For a \emph{pure eigenstate}, $N_{\text{eff}} = 1$, and the DOS-QPE output is that of standard QPE, i.e. a sharp peak centered at that eigenvalue.
\end{itemize}

The effective sampling variance thus scales with the number of contributing eigenstates. Assuming uniform $p_j = 1/N_{\text{eff}}$, the variance of the spectral histogram is
\begin{equation}
    \mathrm{Var}[P(y)] \sim \frac{N_{\text{eff}}}{M \cdot 2^m},
\end{equation}
and the total error scales as
\begin{equation}
    \|\hat{\rho}_S - \rho_S\|_2 \sim \sqrt{ \frac{N_{\text{eff}} \cdot 2^m}{M} }.
\end{equation}
Hence, to achieve error $\delta$, the number of measurements must satisfy
\begin{equation}
    M \gtrsim \frac{N_{\text{eff}} \cdot 2^m}{\delta^2}.
\end{equation}

This highlights the trade-off between resolution, state support, and sampling complexity. Choosing input states with reduced spectral support (e.g., Dicke states) allows one to probe restricted subspaces of the spectrum efficiently.


\section{Rescaling the Hamiltonian}
\label{sec:rescaling_appendix}

For DOS-QPE, the input Hamiltonian must have a spectrum that is real, non-negative, and bounded in the interval $[0, 1)$. Suppose we are given a Hamiltonian $H$ with real eigenvalues $\lambda_i \in [\lambda_{\text{min}}, \lambda_{\text{max}}]$. We construct a rescaled Hamiltonian $\tilde{H}$ such that its eigenvalues lie in $[0, 1)$ as follows:
\begin{equation}
    \tilde{H} = \frac{H - \lambda_{\text{min}} I}{\Delta} ,
\end{equation}
where $\Delta = \lambda_{\text{max}} - \lambda_{\text{min}}$. The resulting eigenvalues are
\begin{equation}
    \tilde{\lambda}_i = \frac{\lambda_i - \lambda_{\text{min}}}{\Delta} \in [0, 1) .
\end{equation}
This procedure requires prior knowledge of both $\lambda_{\text{min}}$ and $\lambda_{\text{max}}$.

The unitary evolution generated by the rescaled Hamiltonian is
\begin{equation}
    \tilde{U} = \exp(2\pi i \tilde{H}) = \exp\left( \frac{2\pi i}{\Delta}(H - \lambda_{\text{min}} I) \right) .
\end{equation}
Running DOS-QPE with $\tilde{U}$ returns phases $\tilde{\lambda}_i$ from which the original spectrum can be recovered via
\begin{equation}
    \lambda_i = \lambda_{\text{min}} + \Delta \tilde{\lambda}_i .
\end{equation}

To simulate $\tilde{U}$ via Trotterization, we decompose $\tilde{H}$ into local terms $\tilde{H}_j$ and approximate the unitary evolution by
\begin{equation}
    \tilde{U} \approx \left[ \prod_j \exp\left( 2\pi i \frac{\Delta t}{r} \tilde{H}_j \right) \right]^r ,
\end{equation}
where $r$ is the Trotter order and $\Delta t$ is the total evolution time. For first-order Trotterization, this becomes
\begin{equation}
    \tilde{U} \approx \exp\left( 2\pi i \frac{\Delta t}{\Delta} H \right) \cdot \exp\left( -2\pi i \frac{\Delta t}{\Delta} \lambda_{\text{min}} \right) .
\end{equation}
The first term corresponds to evolution under the original Hamiltonian $H$ with a rescaled time $\Delta t / \Delta$, while the second is a global phase that factors out in DOS-QPE. However, this global phase must be accounted for when converting the measured phases back to the original eigenvalues.


\bibliography{main}

\end{document}